\documentclass{PoS}
\usepackage{wrapfig}

\title{German-Russian Astroparticle Data Life Cycle Initiative}

\ShortTitle{GRADLC Initiative}

\author{
\speaker{Andreas~Haungs}~$^{1}$, 
Igor~Bychkov~$^{2,3}$, 
Julia~Dubenskaya~$^{4}$, 
Oleg~Fedorov~$^{5}$, 
Andreas~Heiss~$^{6}$,
Donghwa~Kang~$^{1}$,
Yulia~Kazarina~$^{5}$, 
Elena~Korosteleva~$^{4}$, 
Dmitriy~Kostunin~$^{1,7}$,
Alexander~Kryukov~$^{4}$,
Andrey~Mikhailov~$^{2}$, 
Minh-Duc~Nguyen~$^{4}$,   
Frank~Polgart~$^{1}$,
Stanislav~Polyakov~$^{4}$, 
Evgeny~Postnikov~$^{4}$, 
Alexey~Shigarov~$^{2,3}$,
Dmitry~Shipilov~$^{5}$, 
Achim~Streit~$^{6}$,
Victoria~Tokareva~$^{1}$,
Doris~Wochele~$^{1}$,
J\"urgen~Wochele~$^{1}$,
Dmitry~Zhurov~$^{5}$ \\
{$^{1}$} Karlsruhe Institute of Technology, IKP, 76021 Karlsruhe, Germany \\ 
{$^{2}$} Matrosov Inst. f. System Dynamics and Control Theory, Irkutsk 664033, Russia \\
{$^{3}$} Irkutsk State University, Irkutsk 664003, Russia \\
{$^{4}$} Lomonosov Moscow State University, SINP, Moscow 119991, Russia \\
{$^{5}$} Irkutsk State University, Applied Physics Institute, Irkutsk 664003, Russia\\
{$^{6}$} Karlsruhe Institute of Technology, SCC, 76021 Karlsruhe, Germany\\
{$^{7}$} DESY, 15738 Zeuthen, Germany\\
E-mail: \email{andreas.haungs@kit.edu}
}

\abstract{A data life cycle (DLC) is a high-level data processing pipeline that involves data acquisition, 
event reconstruction, data analysis, publication, archiving, and sharing. 
For astroparticle physics a DLC is particularly important due to the geographical and content diversity 
of the research field. A dedicated and experiment spanning analysis and data centre would ensure 
that multi-messenger analyses can be carried out using state-of-the-art methods. 
The German-Russian Astroparticle Data Life Cycle Initiative (GRADLCI) is a joint project of the KASCADE-Grande 
and TAIGA collaborations, aimed at developing a concept and creating a 
DLC prototype that takes into account the data processing features specific for the research field.  
An open science system based on the KASCADE Cosmic Ray Data Centre (KCDC), which is a web-based 
platform to provide the astroparticle physics data for the general public, must also include 
effective methods for distributed data storage algorithms and techniques to allow the community to 
perform simulations and analyses with sophisticated machine learning methods. 
The aim is to achieve more efficient analyses of the data collected in different, 
globally dispersed observatories, as well as a modern education to Big Data Scientist 
in the synergy between basic research and the information society. 
The contribution covers the status and future plans of the initiative. 
}

\FullConference{36th International Cosmic Ray Conference -ICRC2019-\\
		July 24th - August 1st, 2019\\
		Madison, WI, U.S.A.}

\begin{document}

\setcounter{page}{2}

\section{Introduction and Motivation}

\subsection{The High-Energy Universe}

Understanding the high-energy Universe in the context of Astroparticle Physics means first and foremost to answer the urgent question 
of the origin of high-energy cosmic rays. 
Experiments measuring these air-showers could prove that the highest-energy cosmic rays (above ca. 8 EeV) are of extra-galactic origin. 
It is still unknown, however, which sources are responsible for these particles and at which energy exactly the transition from cosmic rays of galactic and extra-galactic origin happens. 
\begin{figure}[b]
\centering
\includegraphics[width=0.5\textwidth]{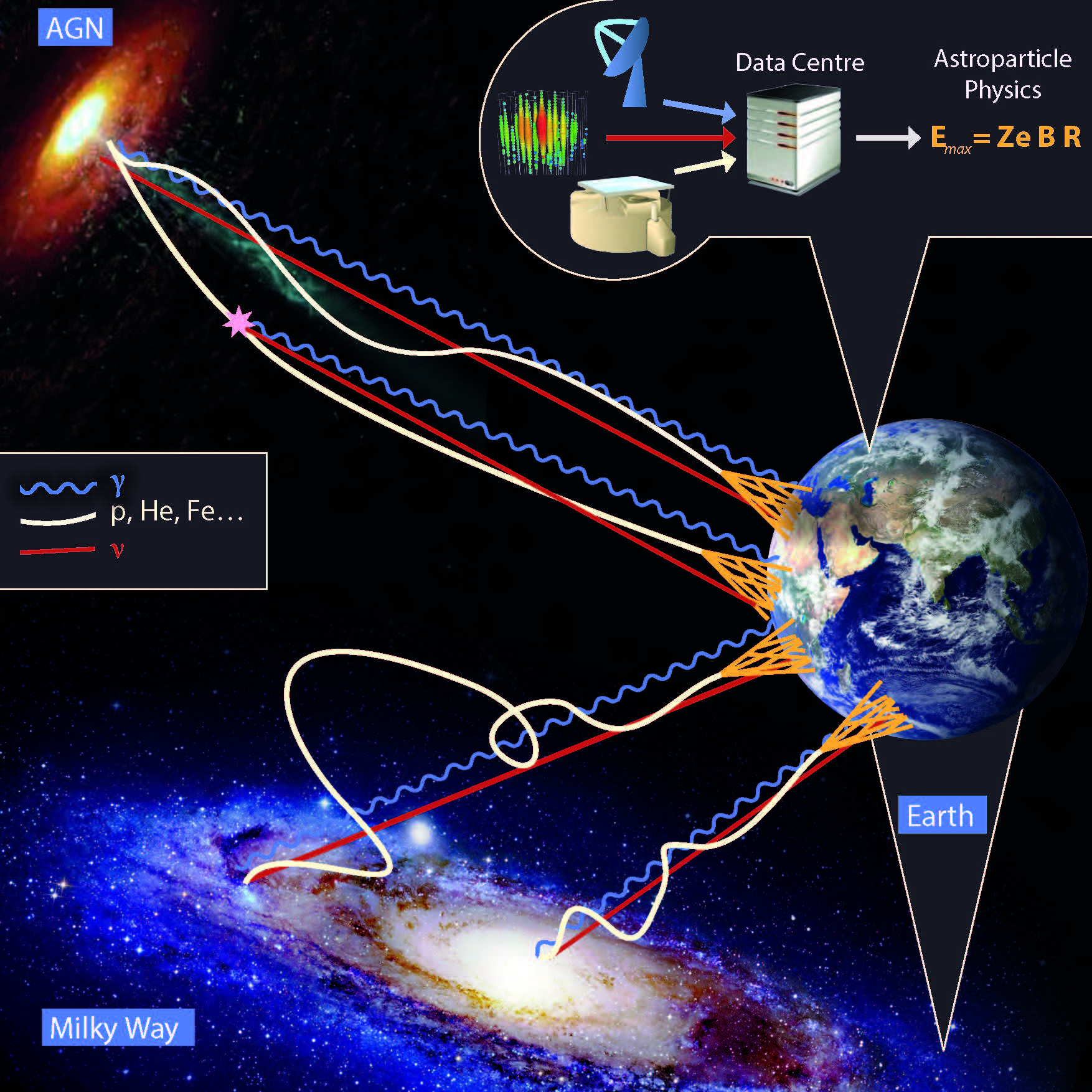} \hspace*{0.2cm}
\includegraphics[width=0.45\textwidth]{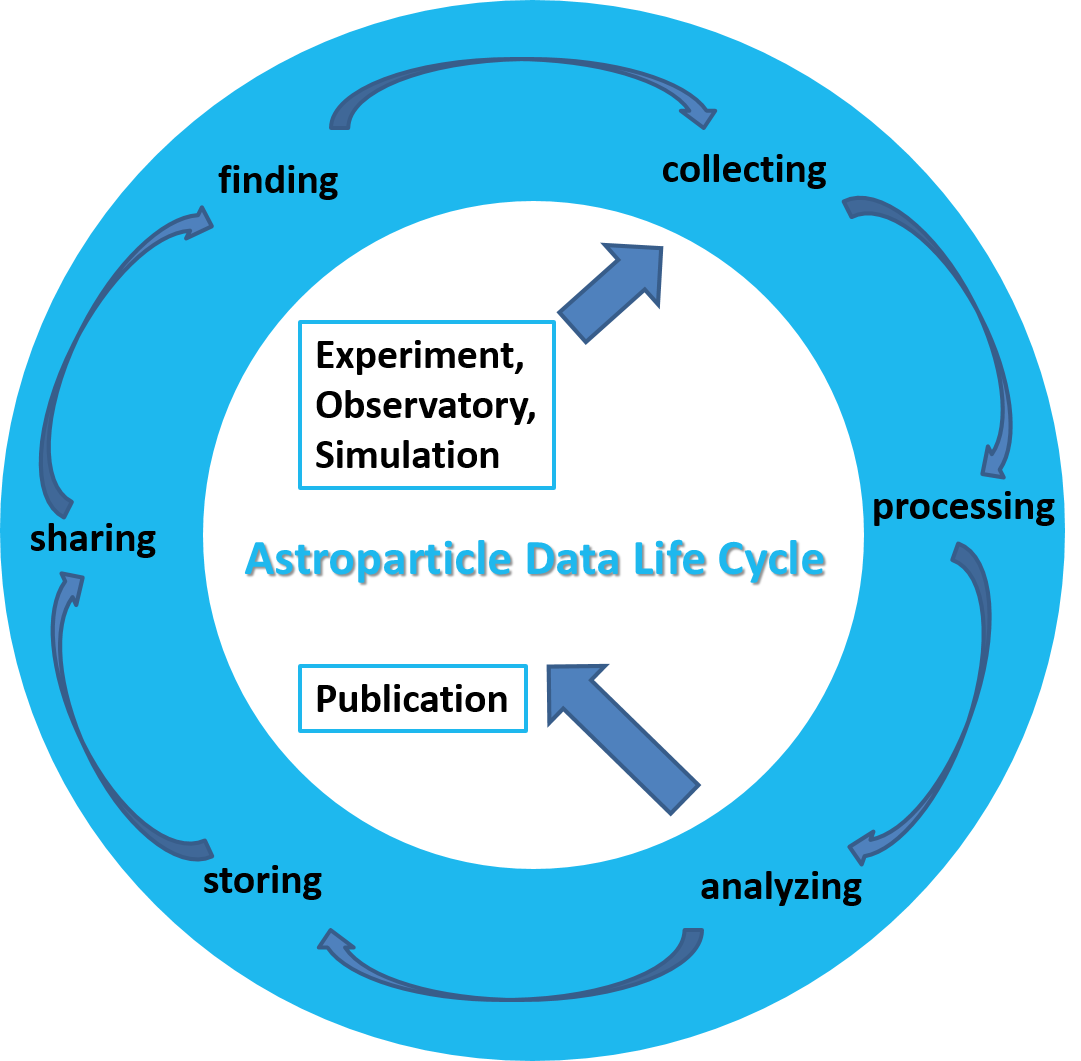}
\caption{Left: Motivation for a global data and analysis centre in astroparticle physics. Cosmic rays, neutrinos and gamma rays of galactic or extra-galactic origin reach our Earth and are observed by different kind of instruments. For  multi-messenger analyses these data have to be combined in a dedicated infrastructure. 
Right: Data life cycle in Astroparticle Physics. Data are generated in the specific observatories or 
experiments and the final goal of each analysis is a (journal) publication of the analysis results. 
A global analysis and data centre has to provide the tools 
and infrastructure to perform each individual part of the cycle.
} \label{fig1}
\end{figure}

Only charged particles can be accelerated in the source regions of the Universe. However, in these acceleration processes also 
secondary neutral products are generated, like gamma-rays and neutrinos. 
Once produced, these particles travel straight on from the source to Earth.
Neutrinos are very weakly interacting with material (detectors) and therefore difficult to measure. 
Gamma-ray measurements are among others motivated by studying them as tracers from charged cosmic-ray sources, but suffer from absorption by the infrared and microwave background. 
Due to all these difficulties it became clear that only a combination of various measurements of the different tracers -- not to forget the rare, but meanwhile detected, catastrophic events in our Universe generating Gravitational Waves -- will bring us closer to an understanding 
of the astrophysics of the High-Energy Universe. 
This approach is called Multi-Messenger Astroparticle Physics (see figure~\ref{fig1}, left panel). 
\begin{figure}[t]
\includegraphics[width=\textwidth]{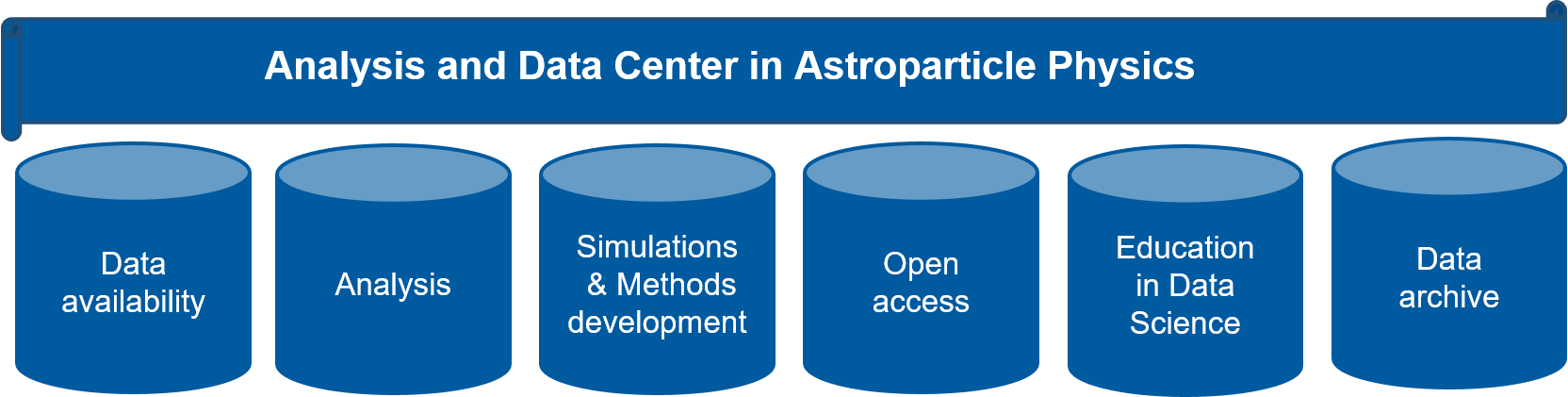}
\caption{The main pillars of a possible global Analysis and Data Centre in Astroparticle Physics:
\textit{Data availability:} All participating researchers of the individual experiments or facilities 
need a fast and simple access to the relevant data.
\textit{Analysis:} A fast access to the Big Data from measurements and simulations is needed.
\textit{Simulations \& Methods development:} To prepare the analyses of the data the researchers need a 
mighty environment on computing power for the production of relevant simulations and the development of 
new methods, e.g. by deep learning.
\textit{Education in Data Science:} The handling of the centre as well as the processing of the data needs 
specialized education in Big Data Science.
\textit{Open access:} It becomes more and more important to provide the scientific data not only to the internal 
research community, but also to the interested public.
\textit{Data archive:} The valuable scientific data need to be preserved for a later (re-)use.
} \label{fig2}
\end{figure}

\subsection{An Analysis and Data Centre for Astroparticle Physics}

With the help of modern Information Technology, Big Data Analytics and 
Research Data Management, we are striving to enter a new era of multi-messenger astroparticle physics. 
This can only be reached if several aspects of a coherent development are considered, where a dedicated 
analysis and data centre is a very important ingredient. It will not only provide the tools and 
environment to take the step into this new physics era, but also allows to take the digitized 
society along the path. 
To efficiently use all the information 
a broad, simple and sustainable access to the scientific data from various infrastructures 
has to be provided. 
In general, such a global data centre must offer a multitude of functionalities that at least cover the pillars shown in the figure~\ref{fig2}.

All the present and future large-scale observatories will provide their scientific data via 
sophisticated infrastructures  and data centres for internal and also external use. 
However, information from various experiments and various messengers like charged particles, gamma-rays 
or neutrinos, measured by different globally distributed large-scale facilities, has to be combined. 
For that a diverse set of astrophysical data is required to be made available and public as well as a framework 
for developing tools and methods to handle the (open and collaborative) data. 
We aim to extend the current activities in the individual observatories on an 
experiment-overarching, global and international level. 
A further goal is to standardize the data and the meta-data, to make the data publication FAIR~\cite{cite25}, 
and by that to make it more attractive for a broader user community. 
The FAIR principles require that all parts of an data life cycle (see figure~\ref{fig1}, right panel) are considered 
equally important in the realisation of an open data centre.  
The move to most modern computing, storage and data access concepts will also open the 
possibility of developing specific analysis methods (e.g.~deep learning) and corresponding simulations 
in one environment opening a new technological opportunity for the entire research field.

\subsection{A First Step towards the Analysis and Data Centre: KCDC} 

Whereas in Astronomy and Particle Physics data centres are already established, which 
fulfill a part of the above mentioned requirements (although, not the same parts), 
in Astroparticle Physics only first attempts are presently under development. 
For example, the KASCADE collaboration~\cite{kascade} has initiated KCDC -- The KASCADE Cosmic-ray Data Centre~\cite{cite21} for a first 
public release of scientific data. In addition, some public IceCube or Auger data can be found 
already now in the Astronomical Virtual Observatories, like in GAVO.

\section{The German-Russian Astroparticle Data Life Cycle Initiative} 

The basics for this common project are delivered by KCDC  and by the operating TAIGA and Tunka-Rex facilities in Russia~\cite{taiga}. 
By many reasons combined Tunka and KASCADE data analyses with sophisticated Big Data Science analysis methods (e.g. deep learning) are of advantage for solving open physics questions. 
These high-statistics experiments are used within GRADLCI as testbed for future multi-messenger astroparticle physics 
analyses based on data of the big observatories coming into operation in next years. 
The project aims, for the first time, for a common data portal of two independent observatories and at the 
same time for a consolidation and maturation of an astroparticle data centre. 

The GRADCLI project~\cite{datajournal} focuses on specific items, all of them are also initial ingredients of the envisaged global data and analysis centre:\\ 
\textbf{Extension of KCDC:} The existing data centre KCDC will be extended by scientific data from 
Russia allowing  on-the-fly multi-messenger analysis. \\
\textbf{Big Data Science Software:} Advancement of Big Data Science: The data centre shall  not only allow access to the data,  but also provide the possibility of developing specific analysis methods and perform corresponding simulations.  The concept to reach this goal is the installation of a dedicated 'Data Life Cycle Lab'. \\
\textbf{Generation of Metadata:} For a FAIR data management it is indispensable to provide a clear concept of meta-data generation and meta-data curation. \\
\textbf{Advanced Data Storage Systems:} A technologically important task is to develop a distributed data storage system, which will allow data of several experiments to be combined into a single repository with a unified interface, as well as will provide the data to all scientists involved. \\
\textbf{Multi-Messenger Data Analysis:} Specific analyses of the data provided by the new data 
centre will be performed to test the entire concept giving important contributions and confidence to the centre as a valuable scientific tool. \\
\textbf{Go for the public:} A coherent outreach of the project, including example applications for all level of users -- from pupils to the directly involved scientists -- with detailed tutorials and documentation is 
an important ingredient of any activity in publishing scientific data. \\
The following are specific examples where GRADLCI made some progress.

\subsection{Data Structure Adaption for Public Re-Use}

Large-scale experiments in astroparticle physics are usually operated several decades by international collaborations of partly several hundreds of scientists. Experiments like KASCADE launched some decades ago, trying to make their data publicly available, suffer from the fact that their data structures cannot be evaluated using modern information technologies.

\begin{wrapfigure}{l}{0.6\textwidth}
\centering
\includegraphics[width=0.6\textwidth]{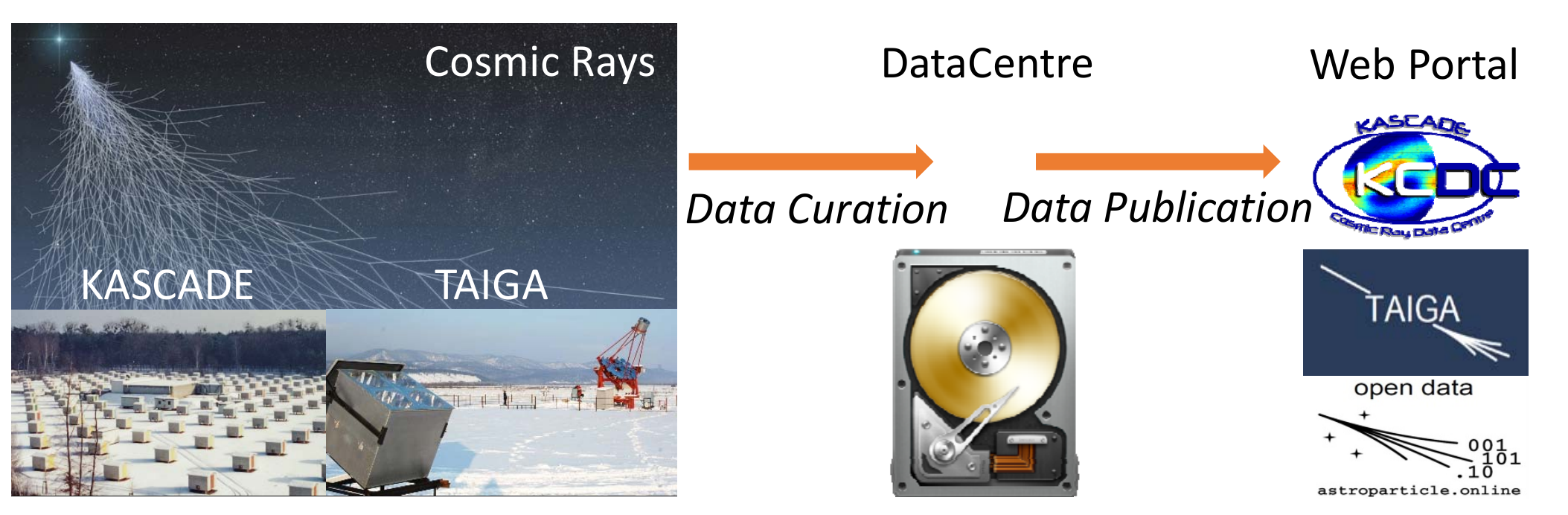}
\caption{Scheme of the data flow of the exemplary data life cycle initiative: Cosmic ray data from two different experiments (KASCADE and Tunka-Rex at the TAIGA site) are stored with a standardized data structure and comparable meta-data before they made available for a re-use.}
\label{Dataflow}
\end{wrapfigure}
To overcome this situation and to guarantee a FAIR (findable -- accessible -- interoperable -- reusable) data preservation, the data must be restructured and reformatted.
A step in this direction is to provide the data and meta-data as well as the tools to analyse the measured data of the meanwhile dismantled cosmic ray experiment KASCADE, which operated from 1996 to 2013. The project to make the entire scientific data public is called KCDC.
The activities within KCDC are used as blueprint for a sustainable data life cycle including aspects of data curation in astroparticle physics.

Published data sets in experimental physics underlie changes whenever analytical methods are improved or errors are discovered. A change of the version number indicates a non-semantic change of the data sets whereas the elements e.g. a specific event, is still part of the data set.
A UUID, implemented in the next release of KCDC, represents such an object and is independent from versions. 
In the next step we will adapt the data of a totally independent experiment to the scheme and include them into KCDC. The software framework to extend KCDC is prepared and in near future the extended KCDC will be released. With a first multi-messenger like analysis applied to the data of both experiments a proof-of-principle of the demonstrator will be given (see figure~\ref{Dataflow}).

\subsection{Meta-Data for FAIR Data Management}

\begin{figure}[b!]
	\centering
		\includegraphics[width=0.48\textwidth]{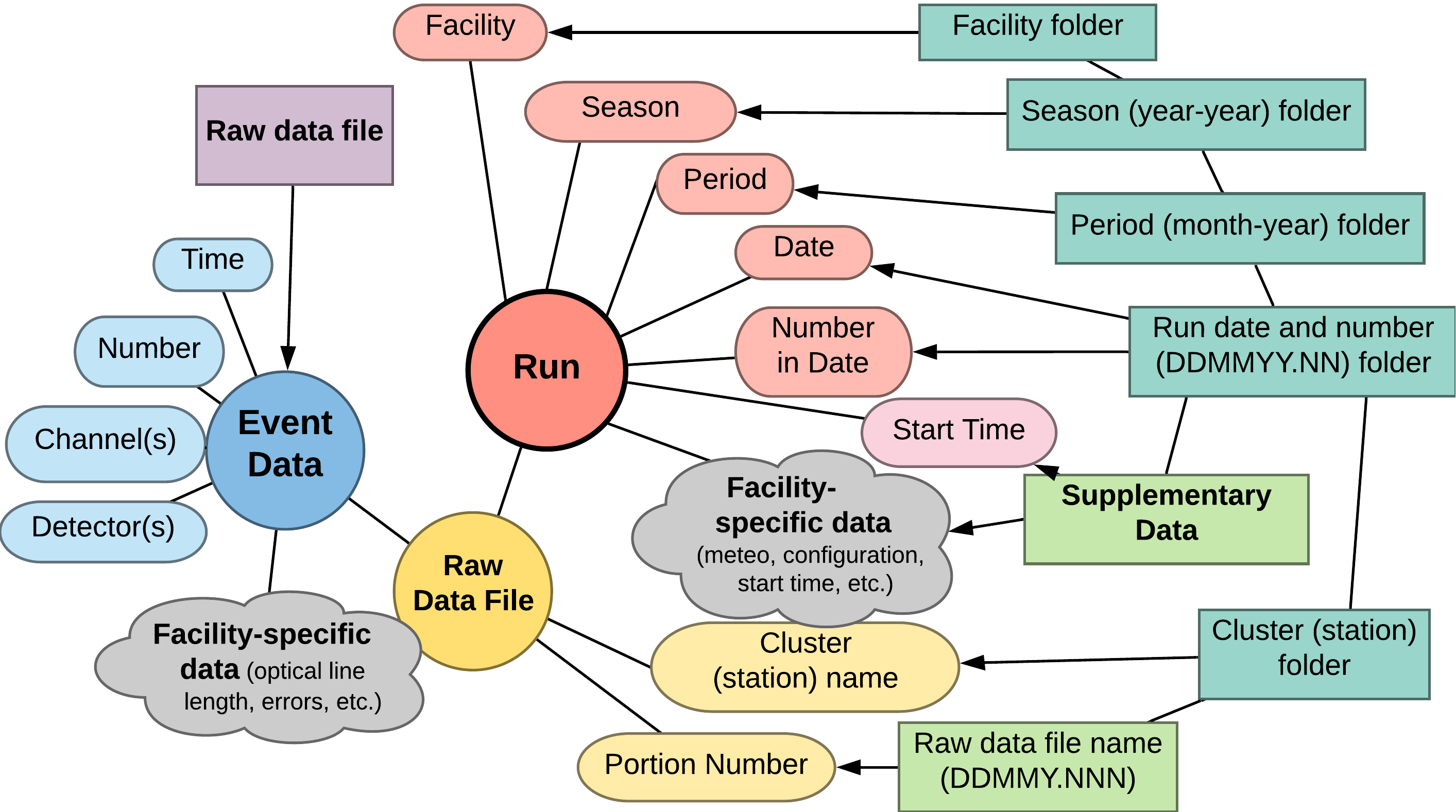} \hspace*{0.2cm}
		\includegraphics[width=0.4\textwidth]{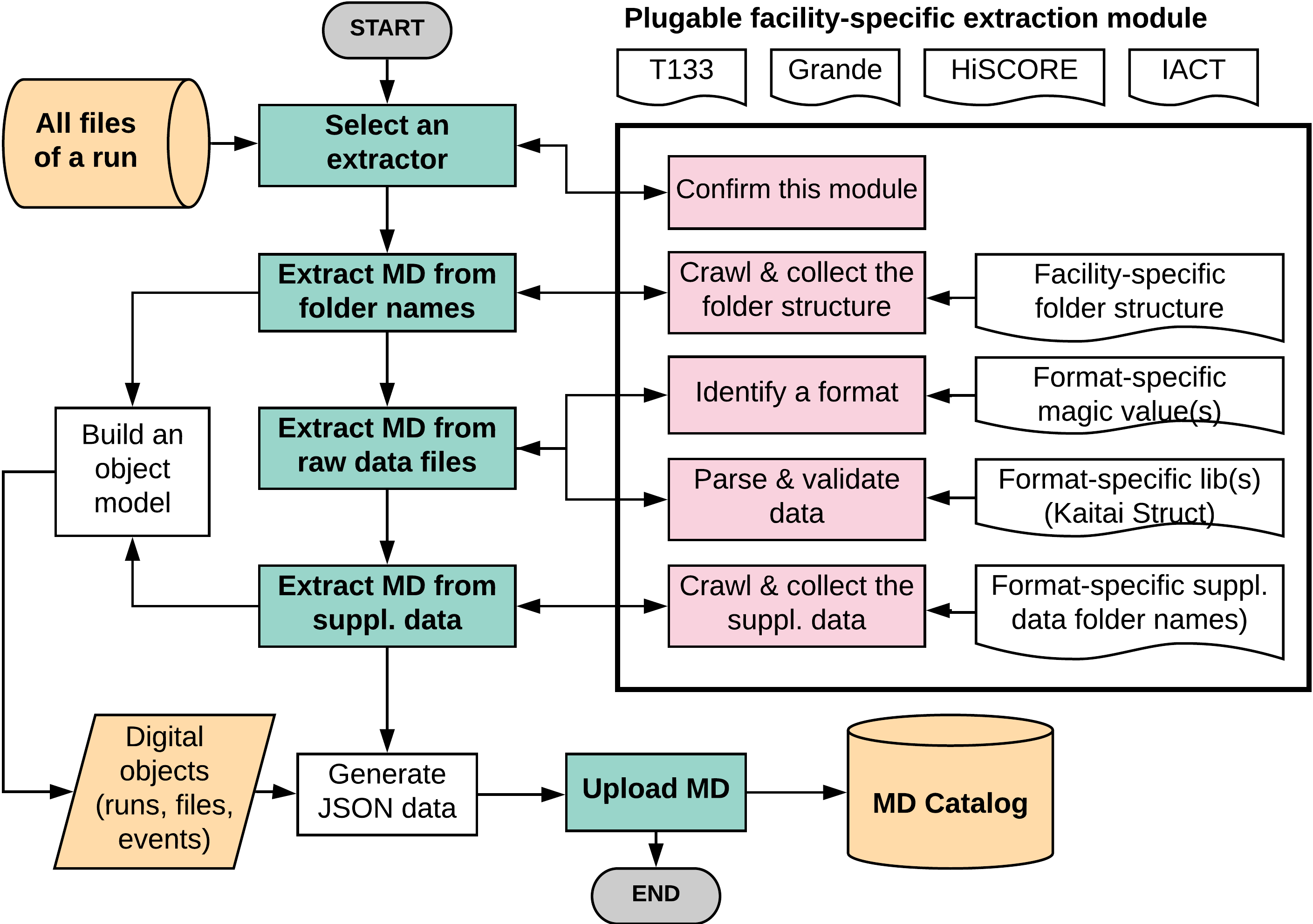}
	\label{fig:md2}
	\caption{Left: General meta-data hidden in TAIGA raw data. Right: Workflow for the meta-data extractor.}
\end{figure}
A coherent concept of providing and generating meta-data is required to ensure a FAIR data management. 
Keeping the raw data ensures facilitating reproducibility of published results and future re-use with a possibly advanced data analysis and processing. 
The TAIGA experiment produces and accumulates a large volume of raw data. 
To make them available for the scientific community they have to be accompanied by meta-data with a unified interface of access. 
Such meta-data should be extracted from binary files, transformed to a unified form of digital objects, and loaded into the catalogue.
To address this challenge we have developed a concept of the meta-data extractor that can be extended by facility-specific extraction modules.
The extractor is aimed to automatically collect descriptive meta-data from raw data files of all TAIGA formats.

Further work for the incorporation of meta-data in the astroparticle data life cycle requires the following steps:
(i) unifying the terminology (conforming a thesaurus); 
(ii) determining a set of user requests to the meta-data catalogue (figure~\ref{fig:md2}, left panel);
(iii) determining a set of hidden and derived attributes describing the digital objects;
(iv) implementing the meta-data extractor (figure~\ref{fig:md2}, right panel);
(v) developing the meta-data catalogue implementing a unified interface of access.

\subsection{APPDS: AstroParticle Physics Distributed Storage}

\begin{wrapfigure}{l}{0.5\textwidth}
\centering
\includegraphics[width=0.48\textwidth]{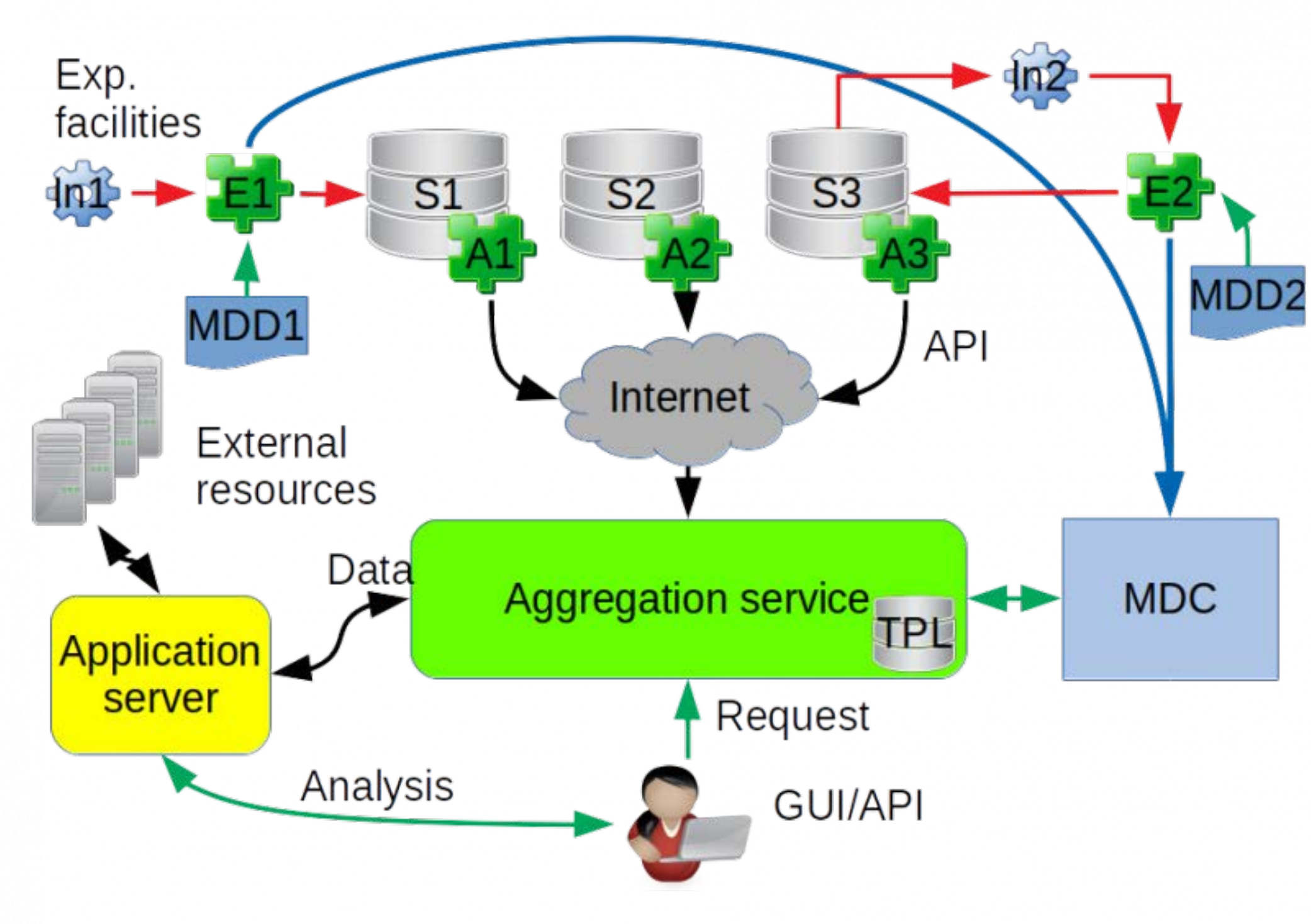}
\caption{Scheme of the APPDS architecture.} \label{figure_appds_architecture}
\end{wrapfigure}
GRADLCI aims to develop a distributed data storage system by the example of the two experiments KASCADE and TAIGA, as well as to demonstrate its viability, stability and efficiency.
An overview of the foreseen architecture of APDDS is presented in Figure~\ref{figure_appds_architecture}. 
S1, S2, S3 are data storage instances of the physical experiments. In1 is the original data input. In2 indicates the case when the original data from In1, which are already stored in a storage instance, are being reprocessed. 
At the level of data storage instances a routine called Extractor (E1) is injected into the pipeline. In most cases, input data are files. After standard processing, the files are passed to the Extractor. The Extractor retrieves meta-data from the files using the meta-data description (MDD) provided by the development groups of the physical experiments, sends the meta-data to the meta-data database using its API, and passes the files back to the pipeline. If the data need to be reprocessed, the same pipeline is applied but with a different type of the Extractor (E2).
The files of each storage instance are delivered to the Data Aggregation Service by the Adapter, which is a wrapper of the CernVM-FS server.
To retrieve necessary files, the user forms a query using the web interface provided by the Data Aggregation Service. When received, it asks the Meta-data Database for the answer and the Data Aggregation Service generates a corresponding response and delivers it to the user.
Finally there is the application server where the distributed computing and the machine learning applications will  happen. 
All components of APPDS are talking to each other via RESTful API.  The key business logic of APPDS is implemented in the Data Aggregation Service whose design and implementation are subject of present work. Some services are running in Docker containers.

\subsection{Workflow Example}

\begin{figure}[ht]
\centering
  \includegraphics[width=0.7\textwidth]{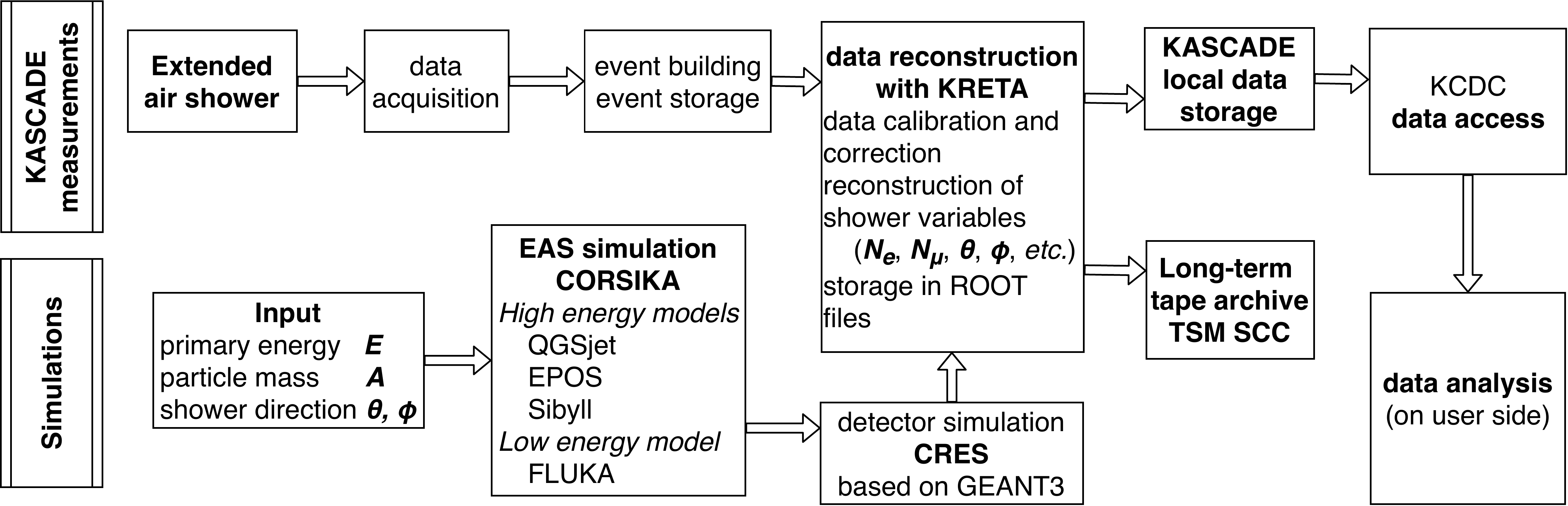}
  \caption{Example of a typical post-operation KASCADE data processing workflow (data life cycle).}
  \label{kascade_wf}
\end{figure}
KASCADE was an experiment at the Karlsruhe Institute of Technology (KIT) in Germany operating from October 1996 till December 2013, corresponding to a total of 4383 days of observation. During this time about 450 million events were collected, which resulted in about 4~TB of reconstructed data.
The data was collected for the purpose to study the spectrum of cosmic rays in the energy range of $10^{14}$--$10^{18}$~eV. There are several levels of reconstructed KASCADE data, starting from the original raw data stored in the CERN ZEBRA format, ending with the high-level of reconstruction shared to the general public.
Data processing is performed by means of special software developed for the experiment: a data reconstruction program KRETA, a program for detector output simulation CRES based on GEANT3 and a program for detailed EAS simulation CORSIKA.
A scheme of the data reconstruction process is presented in figure~\ref{kascade_wf}.
The open access data are stored locally on KASCADE servers and the data storage on magnetic tapes is used as a long-term data preservation. It is employing the Tivoli Storage Manager (TSM) of the Steinbuch Centre for Computing (SCC) at KIT.
For the extension of KCDC as well as for a global analysis and data centre for astroparticle physics, all steps shown in this typical workflow have to be described in detail and made available as services of the data centre.

\subsection{Deep Learning Methods}

Deep learning methods will be more and more used for  data analysis tasks in astroparticle physics. Within our project we aim to make use of the structured access to the data for an efficient application of such sophisticated methods.  
We test the methods within gamma-ray astronomy data analysis by using convolutional neural networks (CNN): 

First, for background rejection (removal of charged cosmic ray events from the data set), the quality strongly depends on the size of learning sample but in any case is substantially better than for conventional techniques~\cite{kryukov5}. 
Another example is the energy estimation of showers from the images in the telescopes, where we achieve a better accuracy than applying conventional approaches for gamma-rays incident in the area outside of a narrow circle around the telescope (100-150$\,$m on ground or 1-1.5 degrees on the camera plane)~\cite{kryukov5}.
We used both PyTorch and TensorFlow toolkits for the development of the optimized CNN. There is still considerable potential to further improve the results by taking into account more accurately the features of the telescopes. 

\subsection{Outreach and Training Aspects}

The outreach and training aspects of GRADLCI are based on KCDC as well as on a new framework developed in Irkutsk: astroparticle.online~. The main objectives of the future Open Laboratory are training experts and developing new instruments and methods for the multi-messenger astronomy as well as supporting open software and open data initiatives.
Taking this and the present environment into account we can define the main pillars of it as:  
 \textit{Development of open training programs}. All programs and their sources (i.e. scripts, slides, problems, etc.) developed in the frame of this laboratory will be published online under free license and can be adopted by the third-party institutes and lecturers. 
These lectures and seminars will be given at ISU (see below) and kept alive and updated. 
\textit{Focus on modern IT and open source solutions}. The modern physics analysis suffers from the lack of the experts in big data and deep learning.
We plan to spend significant efforts on training of these experts during their education at ISU, attracting new experts and trying to keep them in science.
Additionally we will focus on data analysis using modern methods. 
 \textit{Interaction between different facilities}.
The multi-messenger astronomy implies data transfer between astroparticle experiments, 
which can be complicated by the data policies established by the different collaborations.
Within experiments located in the Baikal region we will focus on the policies, exchange protocols and software for the multi-messenger astronomy.

\section{Conclusion}

Several initiatives have been started towards a dedicated and global Analysis and Data Centre in Astroparticle
Physics. The aim of these initiatives, and in particular the GRADLC Initiative, is to develop and implement an interdisciplinary framework, 
which meets the needs of the digitization of the research field and which  is also attractive to society. 
One of the immediate goals is to enable a more efficient analysis of the data that is recorded in 
different locations around the world for coherent multi-messenger studies in order to better understand the high-energy processes in our Universe. 

\noindent
\paragraph{Acknowledgement}
This work was financially supported by Russian Science Foundation Grant 18-41-06003 and the Helmholtz Society Grant HRSF-0027.

\end{document}